 \documentclass[12pt,a4paper]{article}
 \usepackage{epsf}
\begin{document}
\title{Topcolor assisted technicolor models and muon anomalous magnetic moment}
\author{Chongxing Yue$^{a}$, Qingjun Xu$^{b,c}$, Guoli Liu$^{b}$
 \\ {\small a: CCAST (World
 Laboratory) P.O. BOX 8730. B.J. 100080 P.R. China}\\
  \small{and College of Physics and Information Engineering,}\\
 \small{Henan Normal University,
 Xinxiang  453002. P.R.China} \\
{\small b: College of Physics and Information Engineering,}\\
 \small{Henan Normal University,
 Xinxiang  453002. P.R.China}\\
 {\small c: Department of Physics, Zaozhuang Teacher's college,
 ShanDong 277160. P.R.China}
\thanks{E-mail:cxyue@public.xxptt.ha.cn} }
\date{\today}
\maketitle
\begin{abstract}
\hspace{5mm}We discuss and estimate the contributions of the new
particles predicted by topcolor assisted technicolor(TC2) models
to the muon anomalous magnetic moment $a_{\mu}$. Our results show
that the contributions of Pseudo Goldstone bosons are very small
which can be safely ignored. The main contributions come from the
ETC gauge boson $x_{\mu}$ and topcolor gauge boson $Z^{\prime}$.
If we demand that the mass of $Z^{\prime}$ is consistent with
other experimental constrains, its contributions are smaller than
that of $x_{\mu}$. With reasonable values of the parameters in TC2
models, the observed BNL results for $a_{\mu}$ could be explained.
\end {abstract}

\vspace{0.5cm}\noindent {\bf PACS number(s)}: 12.26NZ, 14.60.Ef,
13.40.Em

\newpage
A new experimental value of the muon anomalous magnetic moment,
$a_{\mu}=\frac{1}{2}(g-2)_{\mu}$, was recently reported by
BNL\cite{x1}. The data seems to include a 2.6 standard deviation
from theoretical predictions based on the standard model(SM). This
may be an indication of new physics if it is more than a
statistical fluctuation. If the deviation comes from the new
physics effects, then at $95\%$ CL, $\delta a_{\mu}^{NP}$ must lie
in the range\cite{x2}:
\begin{equation}
215 \times 10^{-11} \leq \delta a_{\mu}^{NP} \leq 637
\times10^{-11}.
\end{equation}

Many authors have considered various possibilities of interpreting
the deviation in models beyond the SM, such as supersymmetry
theories\cite{x3,x4}, technicolor theories\cite{x5}, muon with
substructure in composite models\cite{x6} and models with extra
neutral gauge bosons or fermions\cite{x7}. In this letter, we
estimate the contributions of the new particles (technipions,
top-pions, ETC gauge bosons and extra $U(1)_{Y}$ gauge boson
$Z^{\prime}$) predicted by topcolor assisted technicolor(TC2)
models\cite{x8} to the muon anomalous magnetic moment $a_{\mu}$.
Our results show that, with reasonable values of the parameters in
TC2 models, the deviation could be explained.

To solve the phenomenological difficulties of traditional
technicolor(TC) theory, TC2 models\cite{x8} were proposed by
combing TC interactions with the topcolor interactions for the
third generation at the scale of about 1 TeV. In TC2 models, there
are two kinds of new gauge bosons: (a) ETC gauge bosons, (b) the
topcolor gauge bosons $B_{\mu}^{A}$ and $Z^{\prime}$. Furthermore,
this kind of models predict a number of Pseudo Goldstone bosons
(PGBs) including the technipions in the TC sector and the
top-pions in the topcolor sector. It has been shown\cite{x9} that
all of these new particles can give corrections to the Z-pole
observablities at LEP and SLD, and thus the LEP-SLD precision data
could give constrains on the parameters of TC2 models. In this
letter, we consider the contributions of these new particles to
the muon anomalous magnetic moment $a_{\mu}$ and see whether the
recent measured value of $a_{\mu}$ can give constrains on the
parameters of TC2 models.

Before numerical estimations, we give the couplings of these new
particles to muons. The couplings of PGBs to muons can be written
as\cite{x8,x10}:
\begin{equation}
\frac{im_{\mu}}{\sqrt{2}F_{T}}[\pi_{p}^{0}\bar{\mu}\gamma^{5}\mu+
\bar{\nu}_{\mu
L}\pi_{p}^{+}\mu_{R}+\bar{\mu}_{L}\pi_{p}^{-}\nu_{\mu R}]+
\frac{im_{\mu}}{\sqrt{2}F_{t}}
(\frac{F_{t}}{\upsilon_{w}})[\pi_{t}^{0}\bar{\mu}\gamma^{5}\mu+
\bar{\nu}_{\mu
L}\pi_{t}^{+}\mu_{R}+\bar{\mu}_{L}\pi_{t}^{-}\nu_{\mu R}],
\end{equation}
where $F_{T}$ is the decay constant of the technipion $\pi_{p}$,
$F_{t}\simeq50GeV$ is the decay constant of the top-pion $\pi_{t}$
and $\upsilon_{w}=\upsilon/\sqrt{2}=174GeV$.

For TC2 models, the underlying interactions, topcolor
interactions, are non-universal and therefore do not possess a GIM
mechanism\cite{x11}. When the non-universal interactions are
written in the mass eigen basis, it lead to the flavor changing
(FC) vertices of the new gauge bosons, such as $Z^{\prime}tc$,
$Z^{\prime}\mu e$ and $Z^{\prime}\mu\tau$\cite{x11,x12}. The
couplings of $Z^{\prime}$ to muons , including the $\mu-\tau$
transition, can be written as\cite{x8,x12}:
\begin{equation}
-\frac{1}{2}g_{1}\tan\theta^{\prime}(\bar{\mu}_{L}\gamma^{\mu}\mu_{L}+
2\bar{\mu}_{R}\gamma^{\mu}\mu_{R})-g_{1}[K_{L}(\bar{\mu}_{L}\gamma^{\mu}\tau_{L})+
2K_{R}(\bar{\mu}_{R}\gamma^{\mu}\tau_{R})],
\end{equation}
where $g_{1}$ is the $U(1)_{Y}$ coupling constant at the scale
$\Lambda_{TC}$, $\theta^{\prime}$ is the mixing angle with
$\tan\theta^{\prime}=g_{1}/(2\sqrt{\pi K_{1}})$. $K_{L}$ and
$K_{R}$ are the flavor mixing factors. In the following
estimation, we will assume $|K_{L}|=|K_{R}|=V_{23}$, which is the
matrix element of the lepton mixing matrix V. In order to explain
the atmospheric neutrino results, one needs an almost maximal
mixing between $\nu_{\mu}$ and $\nu_{\tau}$. Thus, we take a large
mixing between the second and third generation leptons, i.e.
$V_{23}=1/\sqrt{2}$.

For TC2 models, the TC interactions play a main role in breaking
the electroweak symmetry. The topcolor interactions also make
small contributions to the electroweak symmetry breaking, and give
rise to the main part of the top quark mass, $(1-\epsilon)m_{t}$
with a model dependent parameter $\epsilon\ll1$. In TC2 models,
ETC interactions are still needed to generate the masses of the
light quarks and leptons, give masses to PGBs, and contribute a
few $GeV$ to $m_{t}$, i.e. $\epsilon m_{t}$. Ordinary fermions
receive the ETC induced masses via the ETC couplings of
technifermions to ordinary fermions. The couplings of the ETC
gauge boson $x_{\mu}$ to muons can be written as:
\begin{equation}
g_{E}(\zeta_{\mu}\bar{L_{L}}\gamma^{\mu}l_{L}+\zeta_{R\mu}\bar{E_{R}}\gamma^{\mu}\mu_{R}),
\end{equation}
where $L_{L}=(N,E)$ and $E_{R}$ represent technifermions, and
$l_{L}=(\nu_{\mu},\mu)_{L}$, $\mu_{R}$ represent the second
generation leptons. Ordinary fermions couple to the techinfermions
via ETC interactions with the coupling constant $g_{E}$,
$\zeta_{\mu}$ and $\zeta_{R\mu}$ are the coefficients of the left-
and the right-handed couplings.

From eq.2 we can see that the PGBs ($\pi_{p}$ and $\pi_{t}$) can
give the one-loop corrections to the muon anomalous magnetic
moment $a_{\mu}$ via the Feynman diagrams shown in Fig.1.
Calculating Fig.1, we can give the corrections of PGBs to
$a_{\mu}$, which can be approximately written as:
\begin{eqnarray}
\delta a_{\mu}^{PGB}&=&\delta a_{\mu}^{\pi_{p}^{0}}+\delta
a_{\mu}^{\pi_{p}^{\pm}}+\delta a_{\mu}^{\pi_{t}^{0}}+\delta
a_{\mu}^{\pi_{t}^{\pm}} \\ \nonumber &\approx &
\frac{G_{F}m_{\mu}^{2}}{4\pi^{2}\sqrt{2}}\{(\frac{m_{\mu}^{2}}{m_{\pi_{p}}^{2}})(\frac{\upsilon}{F_{T}})^{2}
[\ln(\frac{m_{\pi_{p}}^{2}}{m_{\mu}^{2}})-\frac{7}{6}]+(\frac{m_{\mu}^{2}}{m_{\pi_{t}}^{2}})
[\ln(\frac{m_{\pi_{t}}^{2}}{m_{\mu}^{2}})-\frac{7}{6}]\}.
\end{eqnarray}

The contributions of the charged PGB's to $a_{\mu}$ are much
smaller than those of the neutral PGB's, which have been neglected
in above equation. It has been pointed out that the mass of the
lightest technipion $\pi_{p}^{0}$ is about $100GeV$\cite{x10} and
the mass of the top-pion $\pi_{t}^{0}$ is in the range of $200
\sim 400GeV$\cite{x8,x9}. If we take $F_{T}=40GeV$, which
corresponds to the topcolor assisted multiscale technicolor
models\cite{x9}, then we find $\delta a_{\mu}^{PGB}\leq 1.2\times
10^{-12}$. Thus, the corrections of the PGBs to $a_{\mu}$ can be
safely neglected.

The one-loop corrections of the extra $U(1)_{Y}$ gauge boson
$Z^{\prime}$ to the muon anomalous magnetic moment can be divided
into two parts: one part comes from the coupling
$Z^{\prime}\mu\bar{\mu}$ and the other comes from the FC couplings
. The heavier generations are naturally expected to have larger FC
couplings. Then the most main contributions generated by the FC
couplings to $a_{\mu}$ arise from the FC coupling
$Z^{\prime}\mu\bar{\tau}$. The relevant Feynman diagrams are shown
in Fig.2. Using eq.(3), we can estimate the corrections of
$Z^{\prime}$ to $a_{\mu}$, which are given by:
\begin{equation}
\delta a_{\mu}^{\mu\mu}=\frac{\alpha_{e}^{2}}{12\pi
c_{W}^{4}}\frac{m_{\mu}^{2}}{K_{1}M_{Z^{\prime}}^{2}}
+0(\frac{m_{\mu}^{2}}{M_{Z^{\prime}}^{2}}),\hspace{5mm} \delta
a_{\mu}^{\mu\tau}=\frac{2\alpha_{e}V_{23}^{2}}{\pi
c_{W}^{2}}\frac{m_{\mu}m_{\tau}}{M_{Z^{\prime}}^{2}}+
0(\frac{m_{\mu}^{2}}{M_{Z^{\prime}}^{2}},\frac{m_{\tau}^{2}}{M_{Z^{\prime}}^{2}}).
\end{equation}
Using the relation
$K_{1}M_{Z^{\prime}}^{2}\geq\frac{\sqrt{5}\alpha_{e}s_{W}^{2}M_{W}^{2}A}
{8c_{W}^{4}\sqrt{B_{r}^{exp}(\mu\rightarrow 3e)}}$\cite{x12}, we
can estimate the value of $\delta a_{\mu}^{\mu\mu}$, we find
$\delta a_{\mu}^{\mu\mu}<1\times10^{-14}$, which is at least
smaller than the deviation $\delta a_{\mu}$ reported from BNL by
five orders of magnitude. Thus, we can safely neglect the
contributions from the $Z^{\prime}\mu\bar{\mu}$ coupling. However,
the numerical value of $\delta a_{\mu}^{\mu\tau}$ is dependent on
the value of $M_{Z^{\prime}}$, so we can write $\delta
a_{\mu}^{Z^{\prime}}\simeq \delta a_{\mu}^{\mu\tau}$ and we have:
\begin{equation}
\delta a_{\mu}^{Z^{\prime}}\simeq \frac{2\alpha_{e}V_{23}^{2}}{\pi
c_{W}^{2}}\frac{m_{\mu}m_{\tau}}{M_{Z^{\prime}}^{2}}.
\end{equation}

According the idea of TC2 models, the lepton masses are
dynamically generated by ETC interactions. Then the ETC gauge
boson $x_{\mu}$ which is responsible for the mass $m_{\mu}$ should
have contributions to the anomalous magnetic moment $a_{\mu}$ via
the couplings given in eq.(4). The Feynman diagram is similarly to
Fig.2$a$, only with $x_{\mu}$ replacing $Z^{\prime}$, and the
technifermion line replacing the internal muon line. Then the
contributions of $x_{\mu}$ to the anomalous magnetic moment $
a_{\mu}$ can be written as\cite{x2}:
\begin{equation}
\delta a_{\mu}^{ETC}\simeq
\frac{2(2-\gamma)}{2+\gamma}\frac{m_{\mu}^{2}}{M_{x_{\mu}}^{2}}.
\end{equation}
The anomalous dimension $\gamma$ is in the range of $0-2$. In our
calculation, we take $\gamma$ as a free parameter. Thus, for TC2
models, the total correction to $a_{\mu}$ can be written as:
\begin{equation}
\delta a_{\mu} \simeq \delta a_{\mu}^{Z^{\prime}}+\delta
a_{\mu}^{ETC}\simeq \frac{2\alpha_{e}V_{23}^{2}}{\pi
c_{W}^{2}}\frac{m_{\mu}m_{\tau}}{M_{Z^{\prime}}^{2}}+\frac{2(2-\gamma)}{2+\gamma}\frac{m_{\mu}^{2}}{M_{x_{\mu}}^{2}}.
\end{equation}

In Fig.3, we plot the correction $\delta a_{\mu}$ as a function of
the mass $M_{x_{\mu}}$ for $\gamma=1$ and three values of
$M_{Z^{\prime}}$. One can find $\delta a_{\mu}$ is not sensitive
to the value of $M_{Z^{\prime}}$. As long as $1.1TeV \leq
M_{x_{\mu}} \leq 2.2TeV$, TC2 models could explain the 2.6
standard deviation of the muon anomalous magnetic moment over its
SM prediction.

To see the effects of the anomalous dimension $\gamma$ on the muon
anomalous magnetic moment $a_{\mu}$, we plot $\delta a_{\mu}$ as a
function of the parameter $\gamma$ in Fig.4 for
$M_{Z^{\prime}}=2TeV$ and three values of $M_{x_{\mu}}$. From
Fig.4 we can see that the recent experimental value of $a_{\mu}$
give severe bounds on the parameters of TC2 models. For $\gamma =
2 $, the contributions of the ETC gauge boson $ x_{\mu}$ to the
muon anomalous magnetic moment $a_{\mu}$ are zero and the value of
$\delta a_{\mu}$ generated by the topcolor gauge boson
$Z^{\prime}$ is only $1.5\times 10^{-10}$ for
$M_{Z^{\prime}}=2TeV$. In this case, if we want that TC2 models
could explain the deviation $\delta a_{\mu}= a_{\mu}^{exp}-
a_{\mu}^{SM}$, there must be $M_{Z^{\prime}}\approx 550 GeV$ which
is not consistent with other experimental bounds on the
$Z^{\prime}$ mass\cite{x12}. For $\gamma = 0 $, the contributions
of the ETC gauge boson $ x_{\mu}$ are largely enhanced compared
that of $\gamma = 2 $. The $Z^{\prime}$ mass $M_{Z^{\prime}}$ can
be larger than $ 1.5 TeV $, which is consistent with the limits on
$M_{Z^{\prime}}$ from other experiments. However, the mass of the
ETC gauge boson $ x_{\mu}$ must be in the range $ 1.9TeV\leq
M_{x_{\mu}}\leq 3.3 TeV $.

 TC2 models predict a number of new particles, including PGBs
 (technipions, top-pions), ETC gauge bosons and extra $U(1)$ gauge
 boson $Z^{\prime}$. All of these new particles have contributions
 to the anomalous magnetic moment $a_{\mu}$. We find the contributions
 of PGBs are very small which can be safely ignored. If we demand that
 the mass of topcolor gauge boson $Z^{\prime}$ is consistent with other
experimental constrains, its contributions are smaller than that
of the ETC gauge boson $x_{\mu}$. As long as the muon mass is
generated by the ETC interactions or other strong interactions ,
TC2 models could explain the deviation $\delta
a_{\mu}=a_{\mu}^{exp}-a_{\mu}^{SM}$ for reasonable values of the
parameters.

\vspace{1cm}

{\bf ACKNOWLEDGMENT:}This work is supported by the National
Natural Science Foundation of China(I9905004), the Excellent Youth
Foundation of Henan Scientific Committee(9911) and Foundation of
Henan Educational Committee.

\newpage
\vskip 2.0cm
\begin{center}
{\bf Figure captions}
\end{center}
\begin{description}
\item[Fig.1:]Feynman diagrams for the contributions of PGB's in
TC2 models.
\item[Fig.2:]Feynman diagrams for the contributions of extra gauge
boson $Z^{\prime}$ (a) $Z\mu\bar{\mu}$ coupling; (b)
$Z\mu\bar{\tau}$ coupling.
\item[Fig.3:]The total correction $\delta
a_{\mu}$ as a function of the mass $M_{x_{\mu}}$ for $\gamma=1$,
and $M_{Z^{\prime}}=0.5TeV$ (solid line),
$M_{Z^{\prime}}=1TeV$(dotted line) and
$M_{Z^{\prime}}=2.5TeV$(dashed line).
\item[Fig.4:]The total correction $\delta
a_{\mu}$ in TC2 models as a function of the parameter $\gamma$ for
$M_{Z^{\prime}}=2TeV$, and $M_{x_{\mu}}=1TeV$(solid line),
$M_{x_{\mu}}=3TeV$(dotted line) and $M_{x_{\mu}}=5TeV$(dashed
line).
\end{description}

\newpage
\begin{figure}[pt]
\begin{center}
\begin{picture}(250,300)(0,0)
\put(-50,0){\epsfxsize120mm\epsfbox{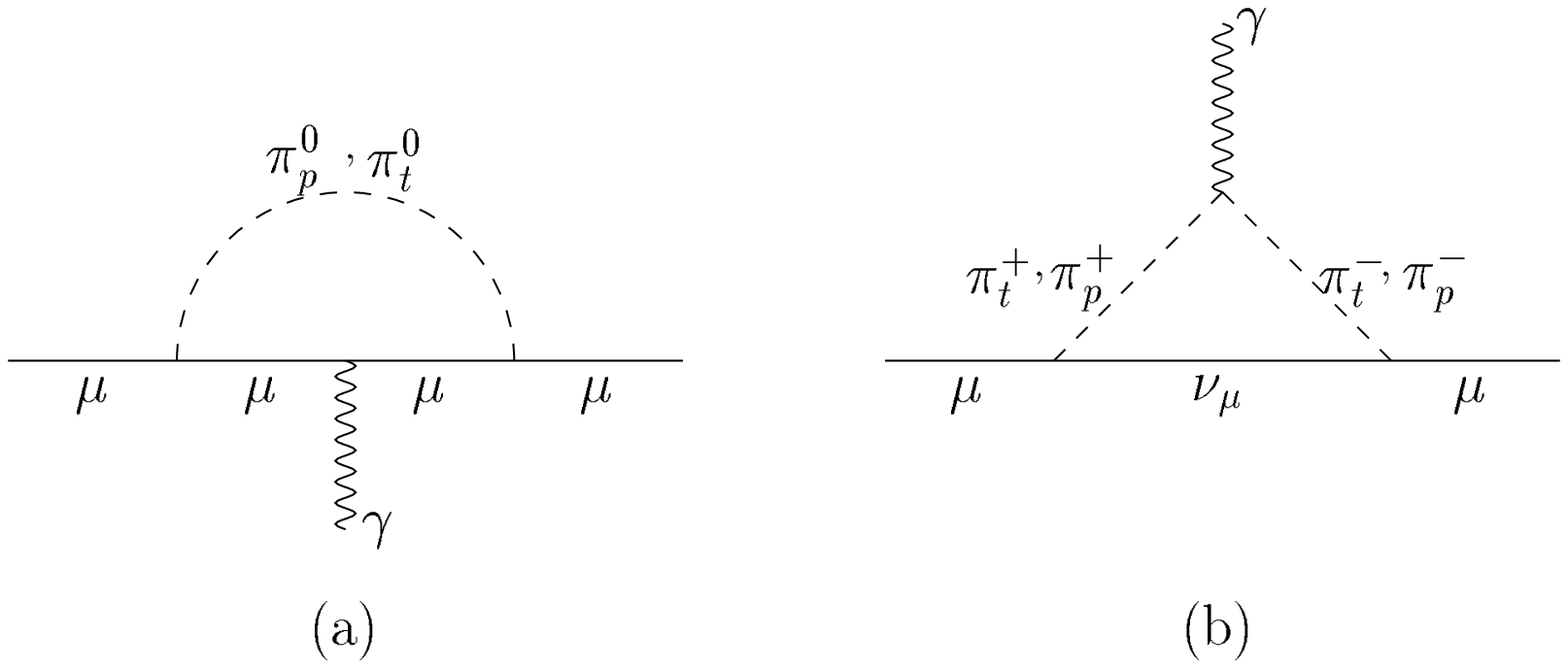}} \put(120,170){Fig.1}
\end{picture}
\end{center}
\end{figure}

\begin{figure}[hb]
\begin{center}
\begin{picture}(250,400)(0,0)
\put(-50,0){\epsfxsize120mm\epsfbox{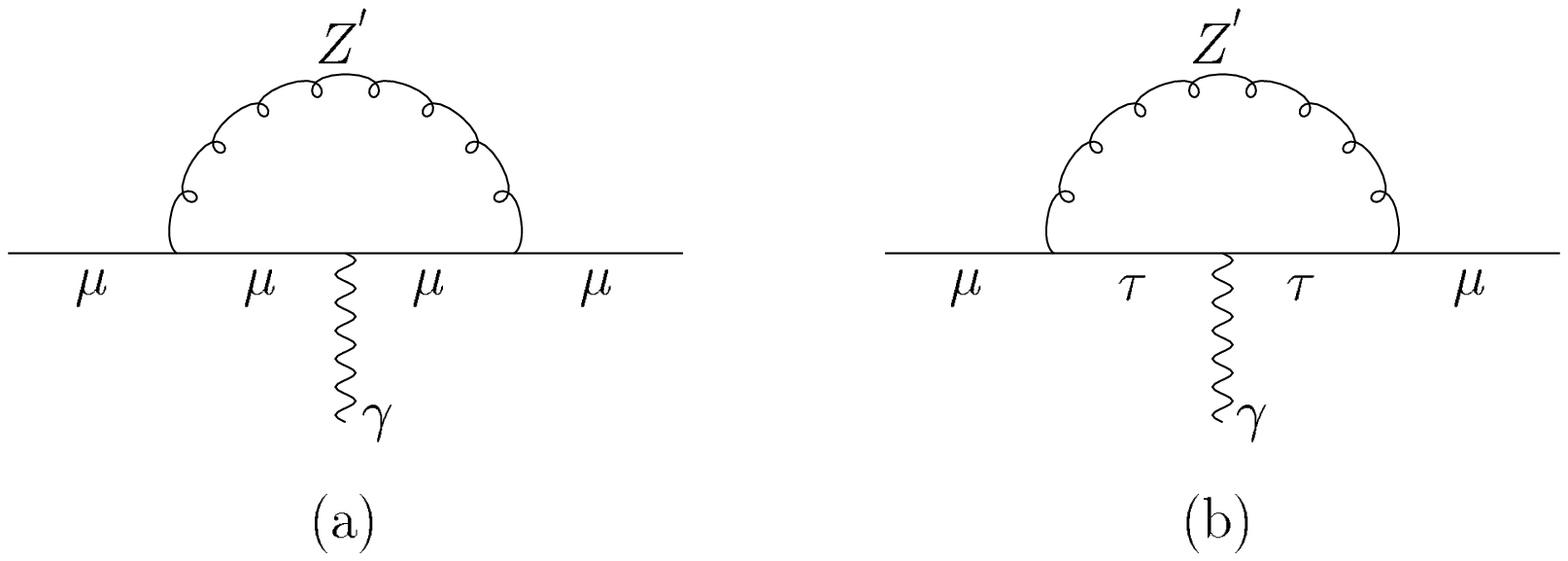}} \put(120,170){Fig.2}
\end{picture}
\end{center}
\end{figure}
\begin{figure}[pt]
\begin{center}
\begin{picture}(250,200)(0,0)
\put(-50,0){\epsfxsize120mm\epsfbox{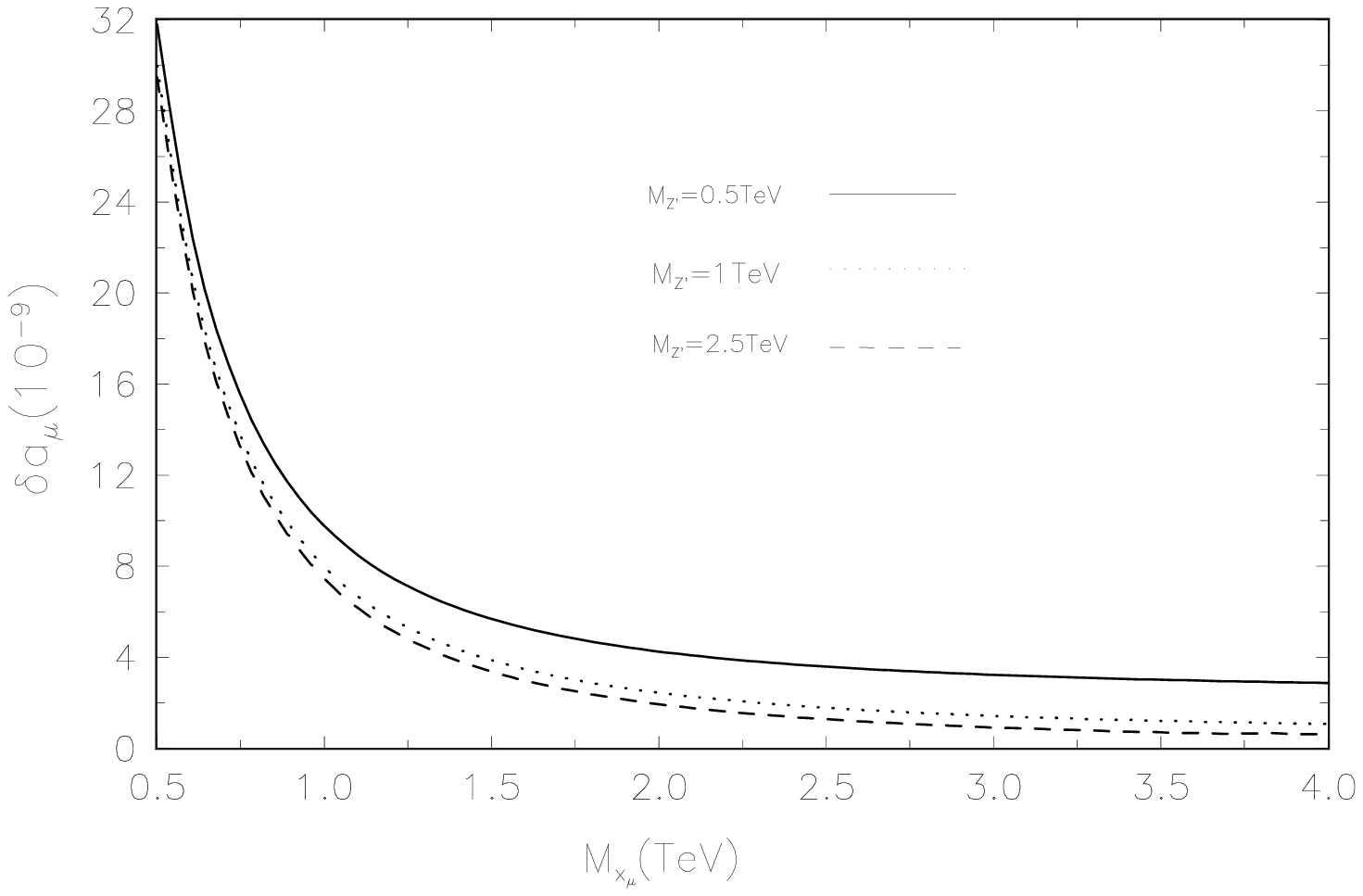}}\put(120,-10){Fig.3}
\end{picture}
\end{center}
\end{figure}
\begin{figure}[hb]
\begin{center}
\begin{picture}(250,200)(0,0)
\put(-50,0){\epsfxsize120mm\epsfbox{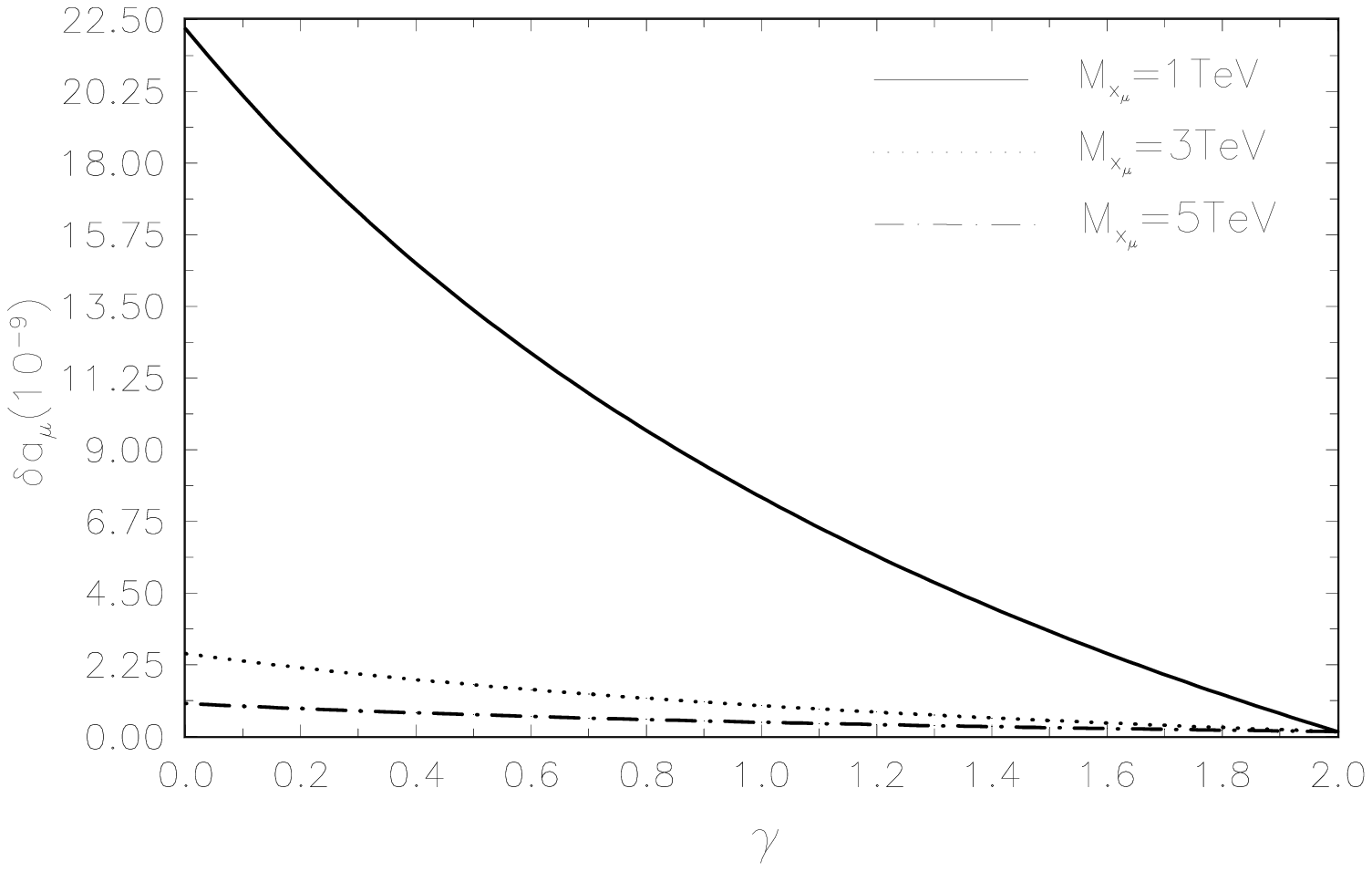}}\put(120,-10){Fig.4}
\end{picture}
\end{center}
\end{figure}
\end{document}